\documentclass[referee]{aa}
\usepackage[bottom]{footmisc}
\usepackage{textgreek}
\usepackage{stfloats}

\usepackage{graphicx}
\usepackage{array}

\usepackage{txfonts}

\begin{document}

   \title{Quantifying the temperature-dependent yields of N$_2$ and N$_2$H$_4$ formation in vacuum-ultraviolet-irradiated NH$_3$ ice}

   \subtitle{}

 \titlerunning{N$_2$ and N$_2$H$_4$ formation in VUV-irradiated NH$_3$ ice}
\authorrunning{Samarth et al.}
   \author{P. Samarth\inst{1,2},
            G. Fedoseev\inst{3},
            S. Ioppolo\inst{4},
            L. Hornekær\inst{4},
            E. F. van Dishoeck\inst{2},
            and
            K.-J. Chuang\inst{1}
            }

\institute{
Laboratory for Astrophysics, Leiden Observatory, Leiden University,
P.O. Box 9513, 2300 RA Leiden, The Netherlands\\
\email{samarth@strw.leidenuniv.nl}
\and
Leiden Observatory, Leiden University,
P.O. Box 9513, 2300 RA Leiden, The Netherlands
\and
Xinjiang Astronomical Observatory, Chinese Academy of Sciences,
Urumqi 830011, China
\and
Center for Interstellar Catalysis, Department of Physics and Astronomy,
Aarhus University, 8000 Aarhus, Denmark
}

\date{Received: 16 January 2026; accepted 18 June 2026}

\abstract
{Gas--grain astrochemical models commonly predict that a substantial fraction of elemental nitrogen in dense cores and protoplanetary disks is locked up in molecular nitrogen (N$_2$), either in the gas or ice phase. However, the direct identification of interstellar N$_2$ ice is still lacking. Measurements by ESA's Rosetta space mission at comet 67P/Churyumov--Gerasimenko revealed a high depletion of N$_2$ with respect to CO, while indicating that NH$_3$ ice is the dominant nitrogen reservoir among the detected species. The origin of the low N$_2$ cometary ice abundance and any relation with NH$_3$ ice is still poorly constrained.}
{This experimental study aims to quantify the conversion of NH$_3$ ice into N$_2$ and N$_2$H$_4$ under various astronomically relevant conditions. With this goal, we systematically investigated how the photochemistry of NH$_3$ ice depends on temperature and photon fluence, which are relevant to star-forming regions.}
{Experiments were performed in an ultra-high-vacuum setup to study the kinetics of the vacuum-ultraviolet (VUV) photon irradiation (115--170~nm) of 40~monolayer-thick NH$_3$ ice at 15, 25, and 70~K up to a total photon fluence of $3.8\times10^{18}$~photons~cm$^{-2}$. Laser desorption with post-ionization reflection time-of-flight mass spectrometry (ReTOF-MS) was applied to monitor N$_2$ and N$_2$H$_4$ formation simultaneously.}
{VUV photolysis of NH$_3$ ice results in the efficient formation of N$_2$ and N$_2$H$_4$, and the formation kinetics and photoproduct ratios depend on temperature and applied photon fluence. The corresponding formation yields were derived from linear fits to the initial product growth, using fluence ranges of $<1\times10^{18}$~photons~cm$^{-2}$ for N$_2$ and $<2\times10^{18}$~photons~cm$^{-2}$ for N$_2$H$_4$. The highest apparent ice-retained N$_2$ formation yield was observed at 15~K, exceeding those at 25 and 70~K by approximately an order of magnitude. The ratio of $\mathrm{N_2}$ to $\mathrm{NH_3}$ column densities was determined as a function of fluence and compared with astronomical constraints. The experiments suggest that NH$_3$ ice photochemistry could represent an additional formation pathway for N$_2$ ice in cold outer planetary environments.}
{}

   \keywords{Astrochemistry --
            Methods: laboratory: solid state --
            ISM: molecules --
            Methods: laboratory: molecular --
            Solid state: volatile --
            Molecular processes}

    \maketitle 

%
\section{\label{sec:level1} Introduction}
Nitrogen is the seventh most abundant element in the Universe. Nitrogen-bearing (N-bearing) species play an essential role in astrochemistry and the origin of life, as elemental nitrogen is a key constituent of amino acids and nucleobases \citep{MunozCaro2002, Bernstein2002, Fedoseev2015, Ioppolo2021, Todd2022LifeFeedstocksNSP}. It has long been suggested that within dense molecular clouds and protoplanetary disks, a large fraction of elemental nitrogen is present in the form of N$_2$ and NH$_3$ \citep{HerbstKlemperer1973DenseClouds, Millar1991UMIST, Herbst2009, Pontoppidan2019}. However, direct observational constraints on solid- and gas-phase interstellar N$_2$ remain limited, as N$_2$ lacks a permanent dipole moment and its far-ultraviolet bands are strongly attenuated by dust, restricting direct searches to diffuse clouds where most nitrogen is still atomic \citep{Li2013A&A}. Instead, N$_2$ abundances are inferred from tracers such as N$_2$H$^+$ and its isotopologs \citep{Womack1992N2Abundances, Daniel2007, Bizzocchi2013, HilyBlant2010, LeGal2014}. Gas--grain astrochemical models constrained by N$_2$H$^+$ observations suggest that solid and gaseous N$_2$ typically account for $\sim$20--50

In contrast, NH$_3$ ice is ubiquitously detected in dense molecular clouds, prestellar cores, and young stellar objects, typically accounting for $\sim$2--10\% relative to H$_2$O ice \citep{Gibb2001, Bottinelli2010c2dIcesIV, McClure2023, Sturm2023}. Besides NH$_3$, only a limited number of N-bearing species have been firmly identified in interstellar ices, primarily NH$_4^+$ and OCN$^-$ \citep{Boogert2015}, while HCN, CH$_3$CN, and related nitriles are constrained by upper limits or tentative detections \citep{Rachid2022, McClure2023, Ozhiganov_2024, Nazari2024CyanidesJWST}. Nevertheless, the combined inventory of identified N-bearing ice constituents accounts for only a small fraction of the total elemental nitrogen budget in dense clouds, with reported values of $\sim$15--30\% \citep{Boogert2015} or $\sim$13\% \citep{McClure2023}. This discrepancy indicates that a substantial fraction of nitrogen (70--85\%) resides in species that are observationally elusive in cold molecular environments, most notably N atoms and N$_2$. Despite indirect evidence for N$_2$ ice in the coldest dense clouds, its abundance remains poorly constrained, with astrochemical models predicting a wide range of N$_2$ ice fractions from 0 to $\sim$35\%, depending on the ice layering and adopted chemical network \citep{FuruyaPersson2018}.

Cometary ices provide an additional perspective on the abundance of N$_2$ ice inherited from earlier evolutionary stages. These ices are thought to preserve a fraction of chemically pristine material from the dense-cloud phase \citep{Mumma2011, Bieler2015}. Mass spectrometric analysis of sublimated cometary material offers an opportunity to directly detect N$_2$ and assess its abundance relative to other volatiles. Using its high-mass resolution, the Rosetta Orbiter Spectrometer for Ion and Neutral Analysis (ROSINA) instrument aboard ESA's Rosetta mission enabled the first secure in situ detections of N$_2$ in comet 67P/Churyumov--Gerasimenko (67P), by resolving N$_2$ from CO. Early mission measurements yielded an N$_2$/CO ratio of $\simeq 5.7\times10^{-3}$, with temporal and seasonal variability leading to increases of up to an order of magnitude near perihelion \citep{Rubin2015, Rubin2019}. Ammonia was also detected, with an NH$_3$/H$_2$O ratio of $\sim$0.1--2\% \citep{BockeleeMorvan2015, Beth2016, Rubin2019}. Observations of other comets, including 1P/Halley and C/1995~O1 (Hale--Bopp), similarly indicate low N$_2$/CO ratios or upper limits \citep{Cochran2000, Altwegg2019, Marty2017}. From the 67P coma abundances, an $\mathrm{N_2}/\mathrm{NH_3}$ ratio of $\simeq 0.13 \pm 0.05$ was derived using the reported N$_2$/H$_2$O and NH$_3$/H$_2$O ratios \citep[Table~2]{Rubin2019}. Notably, both the $\mathrm{N_2}/\mathrm{NH_3}$ and NH$_3$/H$_2$O ratios are about an order of magnitude lower than typical values inferred for dense clouds \citep{FuruyaPersson2018, LeGal2014, Boogert2015, McClure2023}. While the strong depletion of N$_2$ relative to CO can be attributed to the lower trapping efficiency of N$_2$ molecules in the ice \citep{Rubin2015, BarNun2007}, this explanation does not account for the pronounced depletion of NH$_3$. This discrepancy suggests that additional NH$_3$ consumption pathways may operate during cometary ice evolution, potentially involving energetic processing that converts NH$_3$ into N$_2$ and other N-bearing volatiles.

Several formation routes for N$_2$ under astronomically relevant conditions have been proposed, including gas-phase reactions, surface chemistry on cold grains, and energetic processing of ices. Gas-phase reactions of N atoms with abundant neutral species have been invoked as a potential N$_2$ formation pathway in molecular clouds \citep{HasegawaHerbst1993ThreePhase, Herbst2009, LeGal2014}. As cloud density increases, atomic nitrogen accretes onto dust grains, where surface reactions compete between N + N recombination and successive hydrogen addition \citep{Snow2006, FuruyaPersson2018}. Laboratory studies show that hydrogenation dominates under these conditions, efficiently forming NH, NH$_2$, and ultimately NH$_3$ \citep{Hiraoka1995, Watanabe2004, Hidaka2011, Fedoseev2015b}. In contrast, molecular nitrogen is only weakly bound to grain surfaces and thermally desorbs at $\sim$25--28~K, closely following CO desorption and limiting the retention of solid N$_2$ in most cold environments \citep{Bisschop2006DesorptionCON2, Collings2004ThermalDesorptionSurvey, FayolleN2}. In CO-rich ices, N-atom hydrogenation further favors the formation of HNCO \citep{Fedoseev2015}, while NH$_3$ can subsequently participate in acid--base reactions that produce stable ammonium salts at 10--20~K \citep{Raunier2003, Mispelaer2012, Ligterink2018Peptides, Vitorino2024, Slavicinska2025}.

Molecular nitrogen can also be formed through the energetic processing of simple N-bearing molecules in interstellar ice analogs. Ultraviolet (UV) irradiation of pure NH$_3$ ices has been shown to form NH$_2$, N$_2$H$_4$, H$_2$, and N$_2$, accompanied by partial ice-to-gas transfer \citep{Gerakines1996, MartinDomenech2018, Loeffler2010_193nm}. In addition to UV irradiation, radiolysis of pure NH$_3$ by high-energy ions or soft X-rays also suggests the efficient formation of N$_2$ \citep{Chen2009_NEXAFS, Strazzulla2001, Dartois2015, Bergantini2018, Moore2007_review, Loeffler2010_protons}. More recently, \citet{McKinnon2026} showed that photon- and electron-induced processing of NH$_3$-bearing cometary ice analogs can also produce N$_2$. Taken together, these studies demonstrate that energetic processing of NH$_3$ ice can lead to the formation of N$_2$ and other N--N-bearing species over a wide range of irradiation conditions.

Despite this clear evidence, a key limitation of previous work is the lack of quantitative constraints on solid-state N$_2$ kinetics. In situ infrared (IR) spectroscopy cannot be used to track N$_2$ formation because N$_2$ is IR inactive, and most studies therefore rely on quadrupole mass spectrometry (QMS) to identify desorbed N$_2$ either during irradiation or temperature-programmed desorption (TPD). As a result, the absolute yields and formation kinetics of N$_2$ ice as a function of irradiation time or energy dose remain largely unconstrained. This limitation hinders accurate simulations of N$_2$ abundances originating from NH$_3$ energetic processing, obscuring the role of this pathway in the evolution of the elemental nitrogen budget from molecular clouds to planetary systems.

In this work, we employed laser desorption post-ionization reflection time-of-flight mass spectrometry (LDPI--ReTOF-MS) to detect newly formed N-bearing products from the UV photolysis of pure NH$_3$ ice at 15, 25, and 70~K without molecular-specific selection bias. For the first time, the column densities of N$_2$ and N$_2$H$_4$ photoproducts were monitored in situ as a function of photon fluence under controlled laboratory conditions, enabling direct determination of their formation kinetics and yields. The ratio of produced N$_2$ to remaining NH$_3$ was subsequently quantified over the full range of applied VUV photon fluence and substrate temperatures. The experimental setup and procedures are described in Sect.~2, the results and reaction network are presented in Sect.~3, and the astrochemical implications are discussed in Sect.~4.

\section{\label{sec:level2} Experimental methodology}

All experiments were carried out using an ultra-high-vacuum (UHV) cryogenic experimental setup, Mass Analytical Tool to Research Interstellar Ices (MATRI$^{2}$CES), to investigate the evolution of interstellar ice analogs. MATRI$^{2}$CES uses LDPI--ReTOF-MS, to probe the chemical changes in photoprocessed ice samples as a function of VUV irradiation. A detailed description of the setup has been presented in previous studies by \citet{Paardekooper_Bossa_Isokoski_Linnartz_2014} and \citet{Samarth_Bulak_Paardekooper_Chuang_Linnartz_2024}. Here, we only summarize aspects directly relevant to the present work.

\subsection{Experimental setup}

MATRI$^{2}$CES comprises two ultra-high-vacuum (UHV) chambers operating at base pressures in the $10^{-10}$~mbar range: a main reaction chamber and a ReTOF chamber. A gold-plated copper substrate, with dimensions of 16~mm in width and 75~mm in length, is mounted on a closed-cycle helium cryostat. The substrate temperature can be precisely regulated between 15 and 300~K with an accuracy of $\pm$0.5~K, as measured by silicon diodes located at the bottom and top of the substrate, using a resistive heating coil controlled by a temperature controller. Ice samples are prepared on the precooled substrate through vapor deposition introduced by a calibrated precision leak valve with a stainless-steel capillary, which is positioned at $5^{\circ}$ from the surface normal. Ices are irradiated with VUV photons generated by a microwave-discharge hydrogen lamp (MDHL) attached to the main chamber via an MgF$_2$ window. The photon flux at the ice sample position was independently quantified using a NIST-calibrated photodiode, yielding a value of $(2.5 \pm 0.5)\times10^{14}$~photons~cm$^{-2}$~s$^{-1}$, as reported by \citet{Bulak2020}. Its emission spectrum is characterized by Ly$\alpha$ radiation at 121.6~nm (10.2~eV), accompanied by broad H$_2$ Lyman bands spanning in the range of 140--170~nm (7.3--8.9~eV), similar to that reported for a source with identical geometry to the MDHL source by \citet{Ligterink_Paardekooper_Chuang_Both_Cruz_Diaz_Van_Helden_Linnartz_2015, Chen2014, CruzDiaz2014a}. The photon energy distribution is representative of the cosmic-ray-induced UV field in dense molecular clouds \citep{PrasadTarafdar1983}.

The VUV-processed ices are probed with a pulsed Nd:YAG laser (third harmonic, 355~nm, 3--5~ns pulse), focused through a manually adjusted 1~mm aperture onto the substrate to produce a microsecond-scale gas plume of thermally desorbing species \citep{Henderson_Gudipati_2014}, leaving the surrounding ice sample undisturbed. This plume is subsequently ionized by a continuous 70~eV electron beam (Jordan C-950) before interacting with the TOF electrostatic optics and the walls of the setup. The resulting products of dissociative ionization are extracted by a set of electrostatic optics into the ReTOF chamber and detected by a microchannel plate (MCP) detector. Laser firing, extraction voltages, and signal acquisition are synchronized by a Stanford DG-535 delay generator, and ion arrival times are digitized at 2.5~GHz with a dedicated data-acquisition system controlled by a LabVIEW routine. The obtained fragmentation patterns (mass spectra) are further compared with reference mass spectra from databases, such as the National Institute of Standards and Technology (NIST) database \footnote{NIST Chemistry WebBook (2023).}, to identify ice constituents \citep{NIST_ChemWebBook, Bulak_Paardekooper_Fedoseev_Linnartz_2021},  see Appendix~A.

Each measurement cycle comprises 100 laser shots at 5~Hz (see Fig.~1 in \citealt{Samarth2025C2H2Photolysis}). The laser is synchronized with the motorized Y translator to ensure that every pulse probes fresh ice spots within a single column. The 100 spectra are averaged into a single mass spectrum to enhance the signal-to-noise ratio (S/N). The measurement cycle can be repeated on different ice columns by manually shifting the substrate along the X translator in 1.5~mm steps to avoid overlap, yielding a total of eight mass spectra from the same ice sample. This methodology enables probing of ice samples exposed to different cumulative VUV fluences within one control experiment. After each measurement, the substrate is reset to an identical central $XY$ position to ensure reproducible irradiation geometry, while all ion optics are powered down.

\subsection{Experimental procedure}

In this work, the thickness of the ice sample was derived from the deposition time multiplied by the deposition rate, which was predetermined in situ via laser interferometry using a frequency-stabilized He--Ne laser (632.8~nm). The laser beam was incident at $2^{\circ}$ from the surface normal, and the reflected beam intensity was monitored with a photodiode during film growth. The absolute ice thickness, $d$, was calculated from the number of interference fringes observed \citep{Hudgins_Sandford_Allamandola_Tielens_1993}:

\begin{equation}
d = \frac{m \lambda}{2 n_{\mathrm{ice}} \cos \theta_{\mathrm{ice}}},
\label{eq:equation_label01}
\end{equation}

where $m$ is the number of fringes, $\lambda$ is the laser wavelength (632.8~nm), $n_{\mathrm{ice}}$ is the refractive index of NH$_3$ ice, and $\theta_{\mathrm{ice}}$ is the angle of incidence ($2^{\circ}$). For NH$_3$, we adopted $n_{\mathrm{ice}} = 1.34$ \citep{Romanescu2010} and a density of 0.76~g~cm$^{-3}$ \citep{Satorre2013}. The latter allows conversion of absolute thicknesses into column densities. The deposition rate was then obtained by dividing the derived column density by the total deposition time. The estimated thickness of pure NH$_3$ ice (Sigma-Aldrich, $\sim$99.999\%) on the substrate was about $(40 \pm 6)$ monolayers (ML), where 1~ML is defined as $1\times10^{15}$~molecules~cm$^{-2}$.

First, 40~ML-thick pure NH$_3$ ice was deposited at 15~K, corresponding to an astronomically relevant ice thickness of order a few tens of monolayers per icy grain. After deposition, the ice was irradiated for a total period of 256~min with a VUV photon flux of $(2.5 \pm 0.5)\times10^{14}$~photons~cm$^{-2}$~s$^{-1}$. Thus, the NH$_3$ ice was exposed to a total VUV photon fluence of $(3.8 \pm 0.8)\times10^{18}$~photons~cm$^{-2}$. The changes occurring within the pure NH$_3$ ice were tracked by collecting LDPI--ReTOF mass spectra at fixed intervals. These intervals were selected such that each new point corresponded to a twofold increase in total VUV fluence, including an additional measurement at 192~min of VUV irradiation. Subsequently, the irradiation experiments were repeated at substrate temperatures of 25 and 70~K following the same methodology. In these cases, ice deposition was performed at 15~K, after which the samples were warmed to 25 or 70~K at a heating rate of 3~K~min$^{-1}$. The irradiation and acquisition of LDPI--ReTOF mass spectra were then carried out under these fixed elevated-temperature conditions.

For quantitative analysis, NH$_3$, N$_2$, and N$_2$H$_4$ column densities were derived from integrated intensities of selected fragment ions in the LDPI--ReTOF mass spectra. The peak areas were derived by direct numerical integration over fixed $m/z$ windows after baseline subtraction, using the mass signals of NH$_3$ ($m/z = 14$--17), N$_2$ ($m/z = 28$), and N$_2$H$_4$ ($m/z = 29$--32). These fragment ranges were used to identify the species and constrain their contributions to the post-irradiation mass spectra. The measured intensities were corrected using the corresponding 70~eV fragmentation patterns from reference spectra (NIST) to account for minor overlapping parent and photoproduct contributions before deriving molecular column densities. Relative sensitivities between species were then corrected using the 70~eV electron-impact ionization cross sections adopted in this work: 3.036~\AA$^{2}$ for NH$_3$ \citep{NIST_EICS_2004}, 2.508~\AA$^{2}$ for N$_2$ \citep{NIST_EICS_2004}, and 3.760~\AA$^{2}$ for N$_2$H$_4$ \citep{Syage1992}. The absolute scaling was normalized to the independently determined initial NH$_3$ column density from He--Ne interferometry, which set the proportionality constant used to convert the corrected signals into absolute column densities for all species.

\section{\label{sec:level3} Results and discussion}

\subsection{LDPI--ReTOF-MS of pure NH$_3$ ice}

\begin{figure}[htp]
\centering
\includegraphics[width=0.9\linewidth]{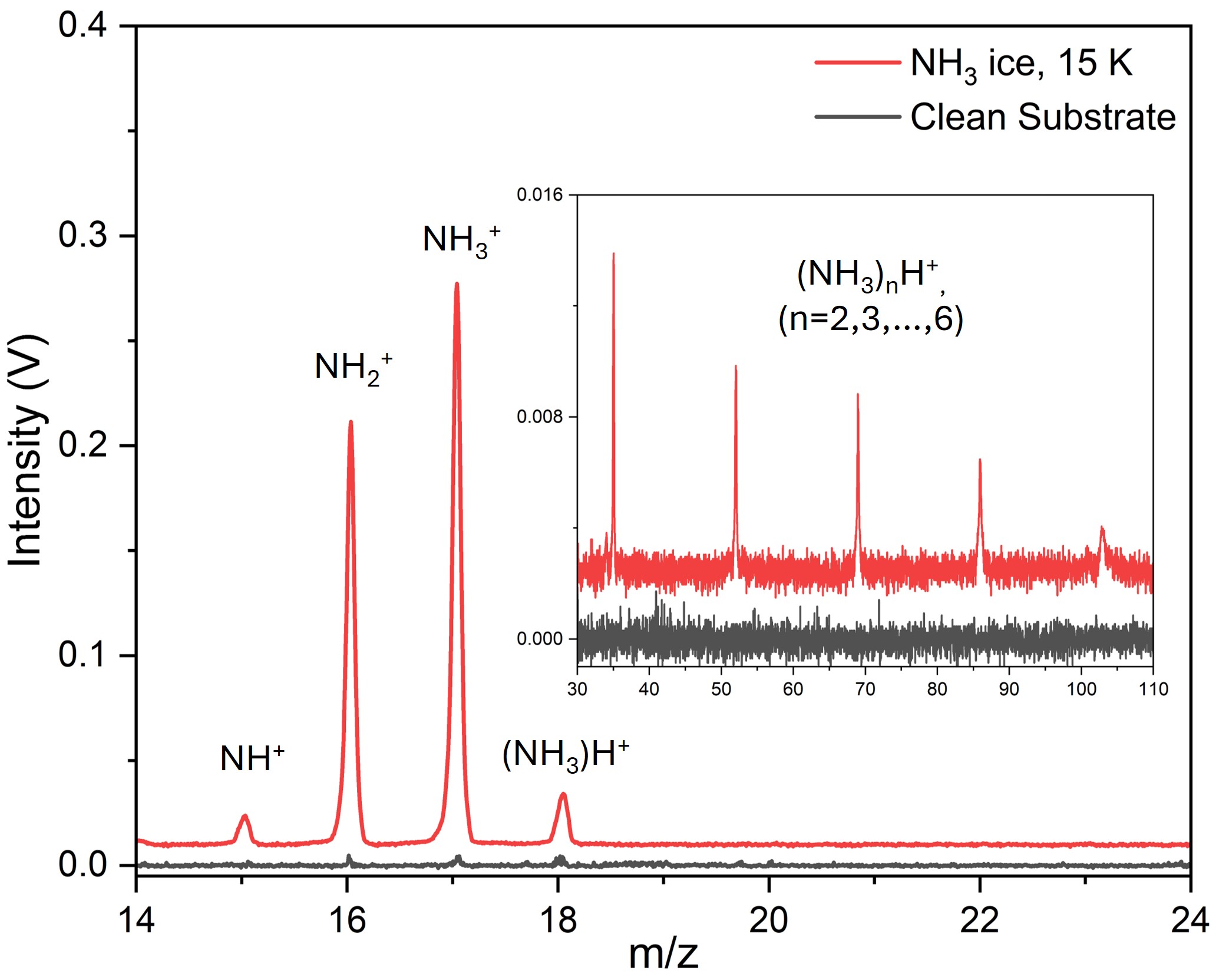}
\caption{LDPI--ReTOF mass spectrum obtained after the deposition of 40~ML pure NH$_3$ ice at 15~K (red), without VUV irradiation. A reference mass spectrum obtained for a bare, clean substrate is presented for comparison (black). The spectra are offset for clarity.}
\label{fig:fig001}
\end{figure}

A representative LDPI--ReTOF mass spectrum of a pure NH$_3$ ice sample at 15~K is shown in Fig.~\ref{fig:fig001} (red). A reference spectrum from a blank experiment obtained under identical experimental conditions is also presented for comparison (black). The obtained spectrum of pure NH$_3$ ice includes major mass signals at $m/z = 15$, 16, and 17. These $m/z$ values correspond to ion fragments following dissociative ionization of NH$_3$ molecules by electron impact at 70~eV, namely NH$^+$, NH$_2$$^{+}$, and NH$_3$$^{+}$. The relative intensities of these three signals match well with the mass fragmentation pattern reported in the NIST database (see Appendix~A). The additional signal at $m/z = 18$ can be assigned to a protonated ammonia molecule, $(\mathrm{NH}_3)\mathrm{H}^+$, which is formed during or subsequent to laser desorption upon 70~eV electron-impact ionization rather than as an intrinsic ammonium component of the ice. This assignment is supported by the observed $m/z = 18$-to-$m/z = 17$ signal ratio ($\sim$0.033), which significantly exceeds the natural nitrogen isotopic ratio ($^{15}$N/$^{14}$N = 0.004). Moreover, the assignment of $m/z = 18$ to $(\mathrm{NH}_3)\mathrm{H}^+$ can be further confirmed by detecting a series of minor signals corresponding to larger protonated ammonia clusters. These signals occur at $m/z = 35$, 52, 69, 86, and 103 and correspond to (NH$_3$)$_n$H$^+$, where $n = 2$, 3, 4, 5, and 6, respectively. The possible contribution of H$_2$O$^+$ to the signal at $m/z = 18$ is considered unlikely. This is due to the absence of (NH$_3$)$_n$(H$_2$O)H$^+$ clusters, for example, at $m/z = 36$ and 53 (see the inset in Fig.~\ref{fig:fig001}). Additionally, no other $m/z$ signals are observed in the range from $m/z = 14$ to $m/z = 110$, indicating limited contamination, such as O$_2$ ($m/z = 32$), CO$_2$ ($m/z = 44$), and N$_2$ ($m/z = 28$), under UHV conditions.

\begin{figure}[!t]
\centering
\includegraphics[width=1\linewidth]{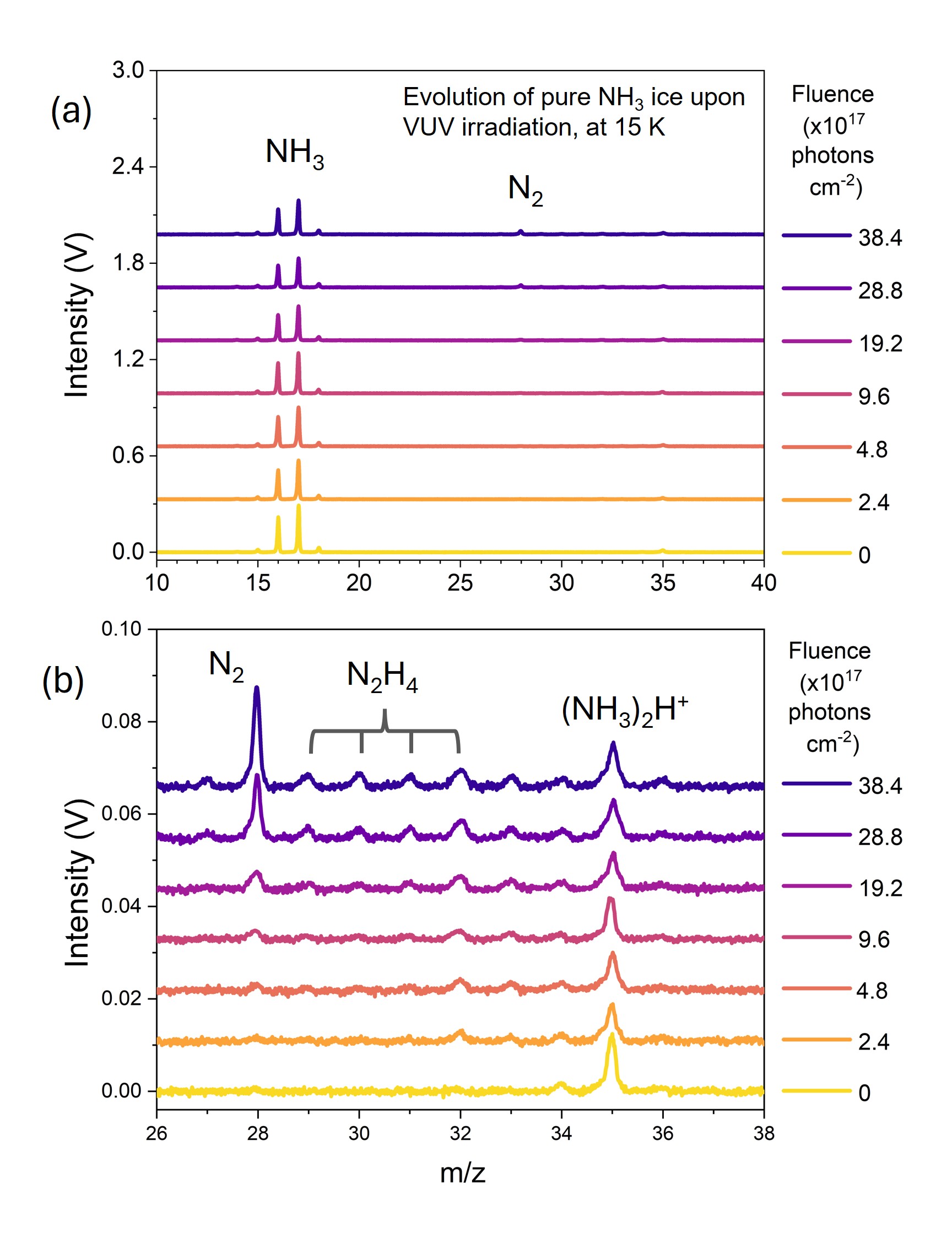}
\caption{Temporal evolution of the LDPI--ReTOF mass spectra obtained during VUV irradiation of a 40~ML pure NH$_3$ ice sample at 15~K. Panel (b) shows the same data as panel (a) on an expanded intensity scale to highlight weaker photoproduct signals. The corresponding photon fluences, ranging from $2.4\times10^{17}$ to $3.84\times10^{18}$~photons~cm$^{-2}$, are indicated next to each spectrum. The unirradiated spectrum is shown for comparison, and spectra are offset for clarity.}
\label{fig:fig002}
\end{figure}

\begin{figure}[htp]
\centering
\includegraphics[width=0.9\linewidth]{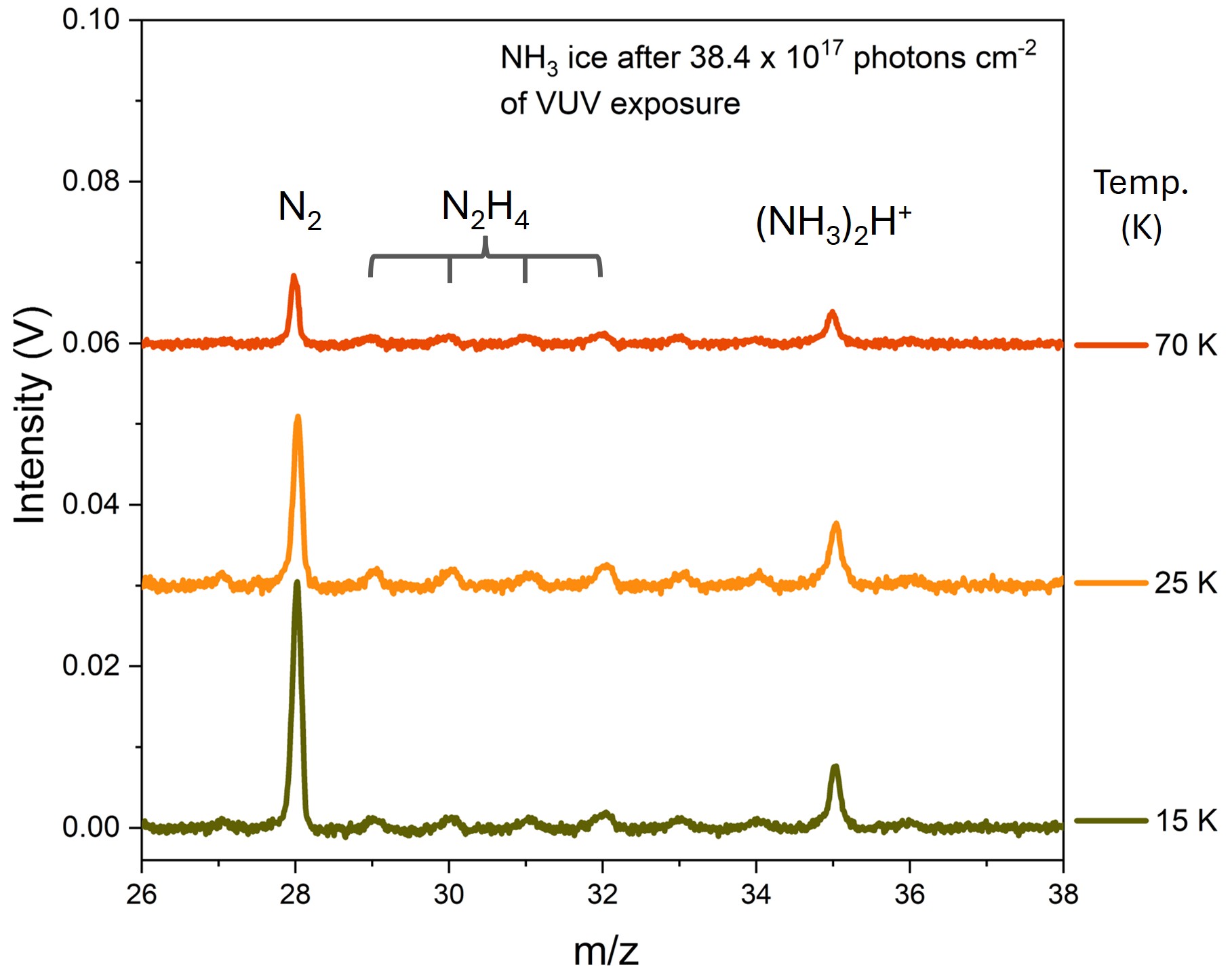}
\caption{Comparison of LDPI--ReTOF mass spectra obtained after VUV irradiation of a 40~ML pure NH$_3$ ice sample at 15, 25, and 70~K with a total photon fluence of $3.8\times10^{18}$~photons~cm$^{-2}$. The spectra are offset for clarity.}
\label{fig:fig003}
\end{figure}

Figure~\ref{fig:fig002} shows a series of time-resolved LDPI--ReTOF mass spectra obtained during VUV irradiation of pure NH$_3$ ice up to a total photon fluence of $3.8\times10^{18}$~photons~cm$^{-2}$ at 15~K. In Fig.~\ref{fig:fig002}a, a selected mass range from $m/z = 10$ to 40 is shown to highlight the evolution of parent NH$_3$ signals. A continuous decrease in the intensity of the $m/z$ signals assigned to NH$_3$ is observed with increasing total photon fluence. The observed depletion of NH$_3$ molecules is attributed to photon-induced reactions and photodesorption. Figure~\ref{fig:fig002}b presents the mass range from $m/z = 26$ to 38 to illustrate the formation of photoproducts during VUV irradiation. The peak at $m/z = 28$ can be assigned to the N$_2$ molecule, as N$_2$$^+$ is the primary mass signal upon electron ionization of N$_2$. In addition, a series of mass signals at $m/z = 29$ (N$_2$H$^+$), 30 (N$_2$H$_2$$^+$), 31 (N$_2$H$_3$$^+$), and 32 (N$_2$H$_4$$^+$) are detected and assigned to N$_2$H$_4$, another expected product of pure NH$_3$ ice irradiation. These peaks exhibit uniform growth with increasing photon fluence. Moreover, the derived relative peak intensities of these four mass fragments agree well with the N$_2$H$_4$ mass fragmentation pattern reported in the NIST database, confirming the assignment to a single species, see Appendix A.

Similar to the aforementioned protonated ammonia cluster peaks at $m/z = 18$ (NH$_{4}$$^{+}$), a peak at $m/z = 33$ can be tentatively assigned to protonated hydrazine ((N$_2$H$_4$)H$^+$). Another possible contribution to the $m/z = 33$ signal is $^{15}$NNH$_4$$^+$. However, the intensity of this peak is too high to be explained by natural isotopic abundances. It should be noted that the mass spectrum of N$_2$H$_4$ reported in the NIST database shows minor peaks at $m/z = 28$ and $m/z = 27$ with relative intensities of 0.21 and 0.003, respectively, with respect to the $m/z = 32$ signal. Therefore, dissociative ionization of N$_2$H$_4$ may contribute to the total counts of the $m/z = 28$ peak assigned to N$_2$. However, this additional contribution is negligible owing to its low fraction and the abundant yield of N$_2$.

The same VUV irradiation experiments were performed at 25 and 70~K to investigate the temperature dependence of the underlying ice chemistry. Figure~\ref{fig:fig003} presents a comparison of the final LDPI--ReTOF mass spectra obtained after VUV photolysis of a pure NH$_3$ ice sample at 15, 25, and 70~K with a total photon fluence of $3.84\times10^{18}$~photons~cm$^{-2}$. The experimental data show very similar photoproducts, including N$_2$ and N$_2$H$_4$, but with distinct absolute abundances. The peak intensity of N$_2$ ($m/z = 28$) exhibits a strong temperature dependence across 15, 25, and 70~K; the observed N$_2$ column density is highest at 15~K and decreases with increasing temperature. The peak intensity of N$_2$H$_4$ is also affected by the ice temperature; however, the highest N$_2$H$_4$ column density is observed at 25~K. To gain deeper insight into the temperature dependence underlying these quantitative differences, the kinetic data for the parent species and photoproducts are analyzed as a function of photon fluence.

\subsection{Analysis of the kinetic data}

\begin{figure}[!t]
\centering
\includegraphics[width=0.7\linewidth]{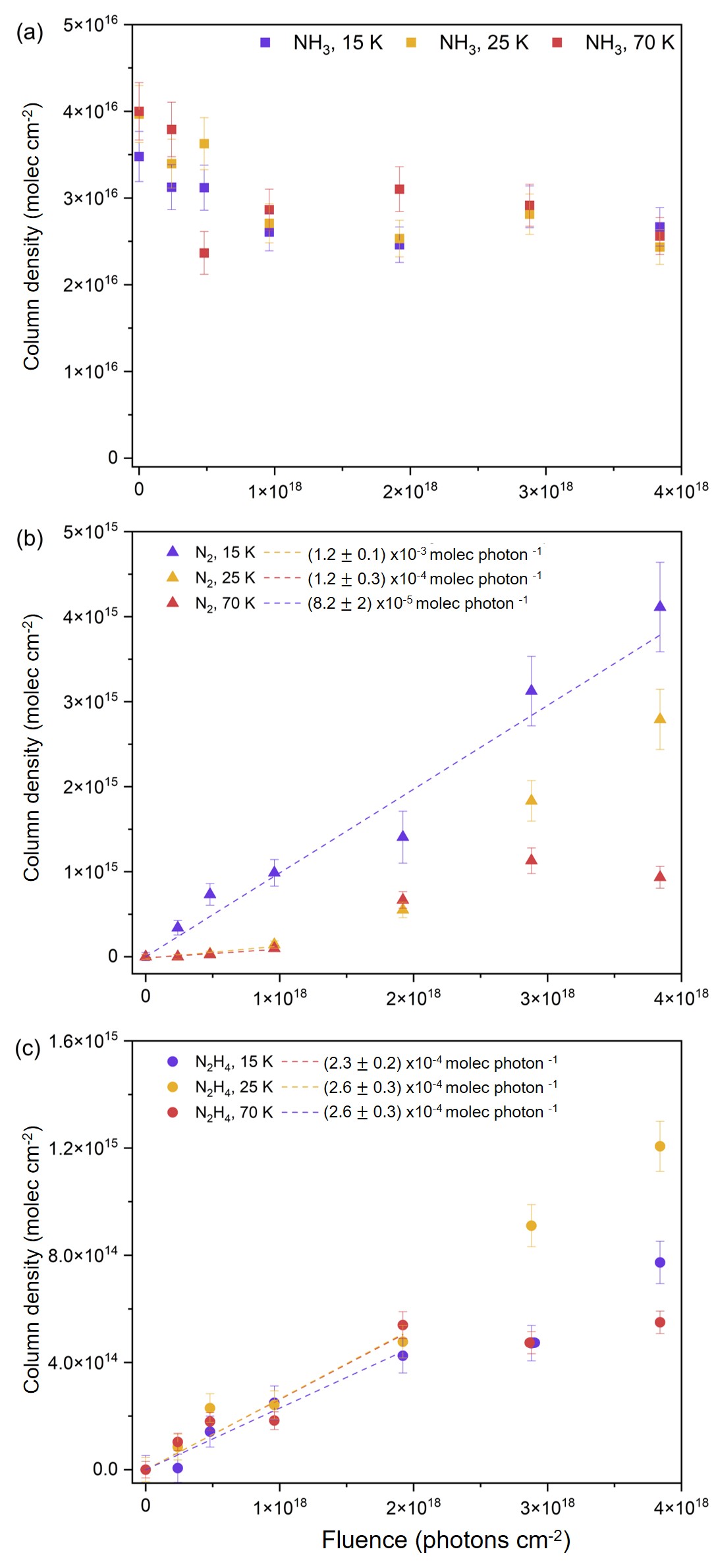}
\caption{Kinetic evolution of NH$_3$, N$_2$, and N$_2$H$_4$ column densities upon VUV irradiation of pure NH$_3$ ice at 15~K (blue), 25~K (yellow), and 70~K (red) as a function of VUV fluence (photons~cm$^{-2}$), shown in panels~(a), (b), and~(c), respectively. Panels~(b) and~(c) additionally show linear fits used to derive the initial apparent formation yields up to a total photon fluence of $1\times10^{18}$~photons~cm$^{-2}$ for N$_2$ and $2\times10^{18}$~photons~cm$^{-2}$ for N$_2$H$_4$. The fitting results (molecules~photon$^{-1}$) are also provided.}
\label{fig:fig004}
\end{figure}

The kinetic curves of the species of interest were obtained by converting the integrated intensities of the observed mass signals into column densities. Details of this conversion are described in Sect.~2.2. The sum of the $m/z = 15$, 16, and~17 signals was used to evaluate the NH$_3$ column density, while the $m/z = 28$ signal was used to determine the N$_2$ column density. The N$_2$H$_4$ column density was derived from the sum of the $m/z = 29$, 30, 31, and~32 signals. The resulting kinetic evolution is presented in Fig.~\ref{fig:fig004} for experiments conducted at 15, 25, and 70~K. Note that the NH$_3$ column density evaluations are relatively scattered during irradiation. Additional uncertainty can arise from changes in the laser-ablation plume profile induced by morphological and compositional alterations of the ice. The measurement repeatability in the applied LDPI--ReTOF-MS quantification of the initial pure ice has been determined to be $\sim$10\%, and can be treated as an instrumental error. However, the realistic uncertainty is expected to be larger, as the ablation and extraction conditions were optimized for the initial pure ice rather than the UV-processed ice samples. The hydrogen-bonded NH$_3$ ice sample produces a more complex plume profile upon ablation compared to simple apolar molecules \citep{Paardekooper2016, Bulak2022}. The plume profile may also be affected by morphological changes in the thick ice during irradiation. Crystallization or compaction of ices upon thermal or energetic processing has been reported previously \citep{JCP2007, Mejia2015, Mifsud2022}. It should be noted that NH$_3$ ice in various astronomical environments is likely mixed with other abundant ice constituents, such as H$_2$O, CO, and CO$_2$. Upon UV irradiation, other solid-state reactions involving NH$_3$ and neighboring molecules are expected to occur. Therefore, the derived solid-state product yields in this study should be regarded as upper limits.

The N$_2$ and N$_2$H$_4$ kinetic curves are reported in Fig.~\ref{fig:fig004}b and Fig.~\ref{fig:fig004}c, respectively. At 15~K, a nearly linear growth of both N$_2$ and N$_2$H$_4$ column densities over the full irradiation period is observed. At 25~K, the N$_2$ kinetic curve demonstrates a distinct behavior compared to that at 15~K. During the initial phase, for a fluence up to $1\times10^{18}$~photons~cm$^{-2}$, the apparent ice-retained N$_2$ formation yield is significantly lower than that at 15~K. At later fluences ($>1\times10^{18}$~photons~cm$^{-2}$), however, the N$_2$ column density increases more rapidly, showing an exponential-like rise. At 25~K, the final N$_2$ column density of $(2.8 \pm 0.4)\times10^{15}$~molecules~cm$^{-2}$ is $\sim$30--40\% lower than that obtained at 15~K. Pure N$_2$ desorption under UHV conditions begins around 25--28~K, whereas N$_2$ retained in polar ice matrices can desorb at somewhat higher temperatures, up to 45~K \citep{FayolleN2, Hodyss2008}. A minor degree of thermal desorption of N$_2$ is therefore expected at 25~K, implying that the observed discrepancy is caused by a difference in the involved reaction rates or routes at slightly elevated temperatures. At 70~K, a much lower N$_2$ yield is detected, reaching a plateau near $(1.0 \pm 0.1)\times10^{15}$~molecules~cm$^{-2}$. This behavior is primarily due to the efficient thermal desorption of newly formed N$_2$ ice.

This evolution of the observed column density reflects the combined effects of solid-state product formation through competing ice chemistry and thermal desorption, both of which are strongly temperature dependent. The N$_2$ kinetic curve suggests a multistep reaction route for N$_2$ formation, in which intermediate species (e.g., N, NH, NH$_2$) gradually accumulate during the early stages of irradiation and are subsequently converted into N$_2$ at later times. The similar N$_2$ column densities observed at 25 and 70~K at fluences $\leq 1\times10^{18}$~photons~cm$^{-2}$ suggest that an additional contribution besides thermal desorption could exist and consequently reduce the N$_2$ abundance at both temperatures. At 25~K, enhanced H-atom mobility can promote rehydrogenation of N-bearing photofragments and thereby suppress the net buildup of N$_2$. Previous solid-state hydrogenation studies show that H-induced reactivity can remain efficient up to $\sim$25~K in systems without a substantial activation barrier \citep{Watanabe2006, Cuppen2010}. At 70~K, H atoms are sufficiently mobile that they do not efficiently participate in subsequent hydrogenation reactions and instead rapidly recombine to form H$_2$, which desorbs into the gas phase. Concurrently, the newly formed N$_2$ also desorbs efficiently at elevated temperatures, thereby reducing its detectable abundance in the solid state.

In contrast to N$_2$, the initial kinetic curves of N$_2$H$_4$, within a fluence of $2\times10^{18}$~photons~cm$^{-2}$, show similar trends at all studied temperatures. The N$_2$H$_4$ curves demonstrate a linear growth with increasing photon fluence. At the end of irradiation, the final column densities are $(0.8 \pm 0.1)\times10^{15}$, $(1.2 \pm 0.1)\times10^{15}$, and $(0.6 \pm 0.1)\times10^{15}$~molecules~cm$^{-2}$ at 15, 25, and 70~K, respectively. The highest abundance is observed at 25~K, about 1.6 and 2.2 times higher than those at 15 and 70~K. At higher fluence, the N$_2$H$_4$ curve, at 70~K shows a clearer leveling-off behavior, whereas the N$_2$H$_4$ curves at 15 and 25~K continue to increase over the measured fluence range.

The final combined column density of N$_2$ and N$_2$H$_4$ in the experiment at 15~K is $\sim(4.8 \pm 0.6)\times10^{15}$~molecules~cm$^{-2}$. Since the formation of one N$_2$ or N$_2$H$_4$ molecule requires two NH$_3$ molecules, this corresponds to $\sim(9.6 \pm 1.2)\times10^{15}$~molecules~cm$^{-2}$ of nitrogen incorporated into the observed solid photoproducts. At 25~K, the final combined column density of N$_2$ and N$_2$H$_4$ is $\sim(4.0 \pm 0.4)\times10^{15}$~molecules~cm$^{-2}$, corresponding to $\sim(8.0 \pm 0.8)\times10^{15}$~molecules~cm$^{-2}$ of nitrogen in the observed products. At 70~K, the final total yield is $\sim(1.6 \pm 0.1)\times10^{15}$~molecules~cm$^{-2}$, corresponding to $\sim(3.2 \pm 0.2)\times10^{15}$~molecules~cm$^{-2}$ of nitrogen in the observed solid products.

To quantitatively compare the initial product growth at various temperatures, a linear fit to the kinetic curves of the products within a fluence of $1\times10^{18}$~photons~cm$^{-2}$ for N$_2$ and $2\times10^{18}$~photons~cm$^{-2}$ for N$_2$H$_4$ was applied to the data shown in Fig.~\ref{fig:fig004}b and Fig.~\ref{fig:fig004}c, respectively. In this work, the reported formation yield is defined as an apparent formation efficiency in units of molecules~photon$^{-1}$, derived from the slopes of the measured solid-state column-density curves as a function of photon fluence. For N$_2$, these values represent the apparent yield in the solid state and do not include N$_2$ that may be formed and subsequently lost from the ice during irradiation. The derived initial formation yields of N$_2$ are $(1.2 \pm 0.1)\times10^{-3}$, $(1.2 \pm 0.3)\times10^{-4}$, and $(8.2 \pm 2.0)\times10^{-5}$~molecules~photon$^{-1}$ at 15, 25, and 70~K, respectively. Thus, in the initial irradiation regime, the apparent ice-retained N$_2$ formation yield at 15~K is about an order of magnitude higher than those at 25 and 70~K, while the 25 and 70~K yields are statistically indistinguishable within the error bars. Thermal desorption of N$_2$ at 25~K cannot be excluded entirely and may contribute to the reduced ice-retained N$_2$ abundance, whereas at 70~K thermal loss is expected to be much more efficient. The observed apparent N$_2$ behavior at 15, 25, and 70~K therefore implies a complex temperature-dependent ice chemistry, including radical diffusion, H-induced reactions, and thermal desorption. In Fig.~\ref{fig:fig004}c, the initial formation yields of N$_2$H$_4$ within a photon fluence of $2\times10^{18}$~photons~cm$^{-2}$ were derived following the same linear fit: $(2.3 \pm 0.2)\times10^{-4}$, $(2.6 \pm 0.3)\times10^{-4}$, and $(2.6 \pm 0.3)\times10^{-4}$~molecules~photon$^{-1}$ at 15, 25, and 70~K, respectively. The nearly constant formation yield suggests a weak or negligible temperature dependence within the uncertainties in the selected fluence range.

\subsection{Possible involved reaction routes}

A more detailed interpretation of the obtained results is not possible without full knowledge of all involved reaction routes. Nevertheless, several plausible reactions can be proposed based on our findings. The formation of N$_2$H$_4$ likely proceeds through recombination of NH$_2$ radicals produced by photodissociation of NH$_3$ molecules, as shown in Reactions~(\ref{eq:nh3_photodissociation}) and~(\ref{eq:nh2_recombination}):

{\setlength{\abovedisplayskip}{4pt}
\setlength{\belowdisplayskip}{4pt}
\begin{equation}
\mathrm{NH_3} \xrightarrow{\text{\small $h\nu$}} \mathrm{NH_2} + \mathrm{H},
\label{eq:nh3_photodissociation}
\end{equation}}

{\setlength{\abovedisplayskip}{4pt}
\setlength{\belowdisplayskip}{4pt}
\begin{equation}
\mathrm{NH_2} + \mathrm{NH_2} \rightarrow \mathrm{N_2H_4}.
\label{eq:nh2_recombination}
\end{equation}}

This interpretation is supported by the observation of a nearly linear growth in N$_2$H$_4$ abundance for most of the kinetic curves (see Fig.~\ref{fig:fig004}c), given the efficient photodissociation of NH$_3$. The observed differences in the final abundance of N$_2$H$_4$ at different temperatures are probably due to variations in the mobility and encounter efficiency of NH$_2$ radicals. At 15~K, the limited mobility of NH$_2$ radicals can lead to a lower formation yield of N$_2$H$_4$ via Reaction~(\ref{eq:nh2_recombination}). NH$_2$ radicals may also be reconverted into NH$_3$ through hydrogenation, as shown in Reaction~(\ref{eq:nh2_h_recombination}):

{\setlength{\abovedisplayskip}{4pt}
\setlength{\belowdisplayskip}{4pt}
\begin{equation}
\mathrm{NH_2} + \mathrm{H} \rightarrow \mathrm{NH_3}.
\label{eq:nh2_h_recombination}
\end{equation}}

At 25~K, the increased mobility of H atoms enhances NH$_2$ production through the sequence shown in Reactions~(\ref{eq:nh3_to_nh}) and~(\ref{eq:nh_to_nh2}):

{\setlength{\abovedisplayskip}{4pt}
\setlength{\belowdisplayskip}{4pt}
\begin{equation}
\mathrm{NH_3} \xrightarrow{\text{\small $h\nu$}} \mathrm{NH} + \mathrm{products},
\label{eq:nh3_to_nh}
\end{equation}}

{\setlength{\abovedisplayskip}{4pt}
\setlength{\belowdisplayskip}{4pt}
\begin{equation}
\mathrm{NH} + \mathrm{H} \rightarrow \mathrm{NH_2}.
\label{eq:nh_to_nh2}
\end{equation}}

In the gas phase, photodissociation of NH$_3$ at Ly-$\alpha$ is expected to predominantly produce NH ($\sim$95\%), with only a minor NH$_2$ channel ($\sim$5\%), while NH$_2$ formation becomes dominant at wavelengths longer than $\sim$132~nm, which overlap with a substantial fraction of the present irradiation wavelengths \citep{Heays2017}.

Since the desorption temperature of N$_2$H$_4$ is reported to be in the 140--180~K range under UHV conditions \citep{Zheng2008}, the decrease in N$_2$H$_4$ formation yields observed at 70~K is not simply due to product desorption. It may also be that formation parent molecules desorb, hence their abundance in the ice is lower than at 15~ and 25~K. Possible reactions attributed to the destruction of N$_2$H$_4$ by mobile H atoms include Reaction~(\ref{eq:n2h4_destruction}):

{\setlength{\abovedisplayskip}{4pt}
\setlength{\belowdisplayskip}{4pt}
\begin{equation}
\mathrm{N_2H_4} + \mathrm{H} \rightarrow \mathrm{NH_3} + \mathrm{NH_2}.
\label{eq:n2h4_destruction}
\end{equation}}

Experimental and theoretical studies report an activation barrier for Reaction~(\ref{eq:n2h4_destruction}) in the range 613--1264~K \citep{Vaghjiani1995_H_N2H4, KannoKito2020_H_Abstraction_Hydrazines}. This barrier is sufficient to inhibit this reaction at 15 and 25~K, but allows this pathway to open at 70~K. Although H atoms are not expected to accumulate at 70~K due to rapid desorption, transient H atoms generated by ongoing NH$_3$ photodissociation may still induce prompt H-atom induced reactions before thermalization and escape. Alternative N$_2$H$_4$ formation routes may involve reactions between NH$_3$ and NH radicals \citep{ManteiBair1968, ZetzschStuhl1981} or other excited or nonthermal (``hot'') NH$_n^{*}$ ($n = 1$--3) species. These intermediates can also originate from photodissociation or photoexcitation of NH$_3$ through processes analogous to Reaction~(\ref{eq:nh3_photodissociation}).

The formation of N$_2$ molecules is likely a multistep process, as supported by the analysis of the N$_2$ kinetic curves (see Sect.~3.2). One previously proposed mechanism involves stepwise hydrogen elimination from NH$_3$ under energetic processing, followed by recombination of N atoms into N$_2$ \citep{Gerakines1996}, as shown in Reactions~(\ref{eq:n_atom_formation}) and~(\ref{eq:n_recombination}):

{\setlength{\abovedisplayskip}{4pt}
\setlength{\belowdisplayskip}{4pt}
\begin{equation}
\mathrm{NH_3} \xrightarrow{\text{\small $h\nu$}} \mathrm{N} + \mathrm{products},
\label{eq:n_atom_formation}
\end{equation}}

{\setlength{\abovedisplayskip}{4pt}
\setlength{\belowdisplayskip}{4pt}
\begin{equation}
\mathrm{N} + \mathrm{N} \rightarrow \mathrm{N_2}.
\label{eq:n_recombination}
\end{equation}}

In addition to N+N recombination, the reaction $\mathrm{N}+\mathrm{NH}\rightarrow\mathrm{N_2}+\mathrm{H}$ provides a chemically plausible N$_2$ formation pathway, particularly given the efficient production of NH radicals under Ly-$\alpha$ irradiation, although its contribution in pure NH$_3$ ice is likely limited by the availability of atomic N and competition with rapid hydrogenation at elevated temperatures.
Other possible N$_2$ formation routes involve consecutive dehydrogenation of N$_2$H$_n$ ($n = 1$--4) species under VUV irradiation.

In this work, the observed decrease in N$_2$ formation yields at 25~K is likely due to the increased mobility of H atoms at higher temperatures. Hydrogenation of intermediate NH$_n$ and N$_2$H$_n$ species acts as a reverse pathway that suppresses the H-elimination steps required for N$_2$ production. In mixed ices, this suppression is expected to be further enhanced because radicals produced by photodissociation (e.g., NH) are spatially diluted and no longer formed in close proximity, reducing the probability of radical--radical recombination pathways leading to N$_2$.
This effect results in a non-negligible reduction of the N$_2$ yield at higher temperatures, such as in the 25 and 70~K experiments (see the middle panel of Fig.~\ref{fig:fig004}). Additionally, the significant decrease in the N$_2$ yield at 70~K is controlled by its efficient desorption above 30~K.

\section{\label{sec:level4} Astrophysical implications}

So far, NH$_3$, along with its protonated form NH$_4^{+}$, has been the main carrier of elemental nitrogen among the species detected in interstellar and cometary ices. \citet{Boogert2015} report typical NH$_3$/H$_2$O ice abundances of $\sim$2--10\% in dense cloud and young stellar object environments, while \citet{McClure2023} finds a relative abundance of NH$_3$ (including NH$_4^{+}$) of $\sim$10--12\% with respect to H$_2$O in pristine cloud ices. By contrast, the deduced bulk composition of ices in comet 67P/Churyumov--Gerasimenko indicates an NH$_3$/H$_2$O ratio of only 0.67\% \citep{Rubin2019}. This reveals a significant depletion of NH$_3$ in cometary ices relative to pristine material, hinting at the possible conversion of NH$_3$ into other species during the long evolutionary history of cometary ices, such as N$_2$, ammonium salts, or various amines \citep{MartinDomenech2018, Altwegg2020, Altwegg2022, Chuang2024, Vitorino2024, Slavicinska2025}. Given the widespread presence of the amino functional group (--NH$_2$) in prebiotic compounds, NH$_3$ has been suggested to be a key precursor of interstellar amino acids \citep{MunozCaro2002, Bernstein2002, Oba2022_NatCommun_Nucleobases, Chuang2024}.

\citet{Boogert2015} and \citet{McClure2023} show that the main carriers of the elemental nitrogen budget in pristine clouds remain unidentified. It has long been considered that N$_2$, one of the main products of VUV photodissociation of NH$_3$ ice, may constitute another reservoir of elemental nitrogen \citep{Boogert2015, Herbst2009}. Thus, in this study, we quantified the UV photoproduction of N$_2$ and N$_2$H$_4$, two major products of NH$_3$ ice photoprocessing, over a wide range of applied UV photon fluences ($<3.8\times10^{18}$~photons~cm$^{-2}$) and ice sample temperatures under fully controlled laboratory conditions. This was enabled by the application of the LDPI--ReTOF-MS technique, which provides, for the first time, direct quantification of the IR-inactive N$_2$ yield without the need for a conventional temperature-programmed desorption procedure and enables investigation of the corresponding kinetics at a fixed sample temperature.

In dense molecular clouds, the gas density typically lies in the range$n_{H}\approx10^{4}$--$10^{6}$~cm$^{-3}$, and dust temperatures decrease to 10--15~K. Although the interstellar radiation field (ISRF) is strongly attenuated within prestellar cores, a local UV field arises from cosmic-ray ionization followed by secondary-electron excitation of molecular hydrogen and radiative decay \citep{PrasadTarafdar1983, Heays2017}. The estimated UV photon flux is $10^{3}$--$10^{4}$~photons~cm$^{-2}$~s$^{-1}$, depending on the cosmic-ray ionization rate. The expected cumulative photon fluence reaches up to $\sim10^{17}$--$10^{18}$~photons~cm$^{-2}$ for a typical molecular cloud lifetime of $\sim10^{6}$~yr. The derived initial formation yields of the photoproducts N$_2$ and N$_2$H$_4$ fall well within this photon fluence window ($<1\times10^{18}$~photons~cm$^{-2}$). It is important to note that N$_2$ ice may already be present due to direct gas-phase accretion or in situ N + N $\rightarrow$ N$_2$ formation on grain surfaces and the detected apparent ice-retained N$_2$ formation yield should therefore be considered an additional contribution to any preexisting N$_2$ abundance.

The present experiments were performed on pure NH$_3$ ice, whereas in interstellar environments NH$_3$ is generally diluted within H$_2$O- or CO$_2$-rich matrices. Such dilution is expected to reduce the probability of radical--radical encounters and to limit the mobility of H atoms and NH$_x$ species, thereby affecting both NH$_2$ + NH$_2$ recombination leading to N$_2$H$_4$ and the multistep reaction pathways that produce N$_2$.

\begin{figure}[!t]
\centering
\includegraphics[width=0.95\linewidth]{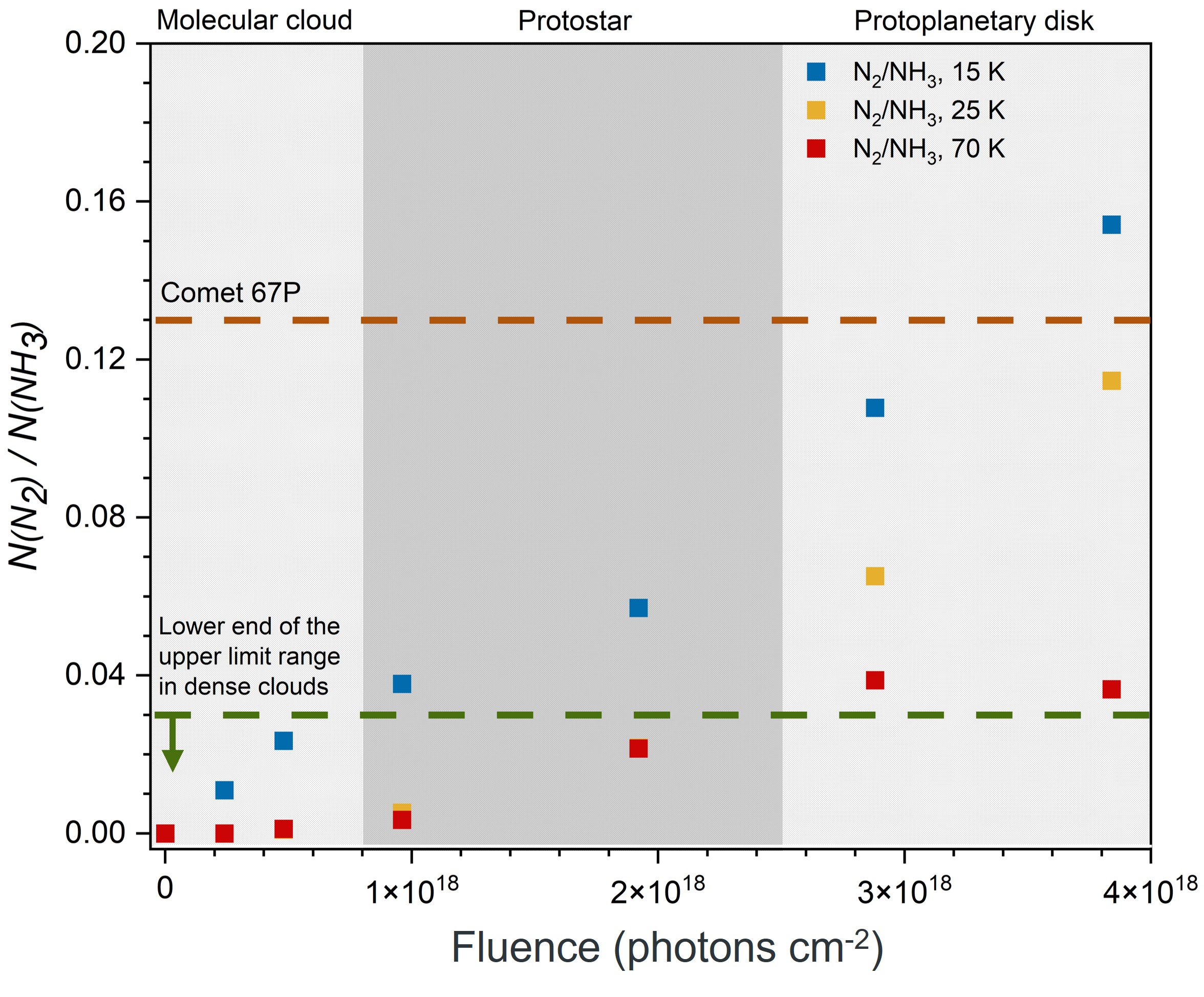}
\caption{Kinetic evolution of the derived $\mathrm{N_2}/\mathrm{NH_3}$ ice ratio as a function of UV fluence. The corresponding UV-fluence ranges for dense clouds, protostars, and protoplanetary disks are shown on the upper $x$-axis. The reported $\mathrm{N_2}/\mathrm{NH_3}$ ratios in observations toward molecular clouds (the lower end of the upper-limit range, green) and comet 67P (orange) are shown as horizontal dashed lines. The different temperatures are indicated by blue, yellow, and red square markers, respectively.}
\label{fig:fig005}
\end{figure}

The kinetic evolution of the derived $\mathrm{N_2}/\mathrm{NH_3}$ ratio as a function of VUV fluence for pure $\mathrm{NH_3}$ photolysis experiments at 15, 25, and 70~K is shown in Fig.~\ref{fig:fig005}. Representative, non-overlapping VUV fluence windows are indicated along the upper $x$-axis: dense molecular clouds ($\sim1$--$8\times10^{17}$~photons~cm$^{-2}$), protostellar envelopes and outer disk surfaces ($\sim0.8$--$2.5\times10^{18}$~photons~cm$^{-2}$), and protoplanetary disk and warm-ice processing histories ($\sim2.5$--$4\times10^{18}$~photons~cm$^{-2}$).

Given the strong chemical link between NH$_3$ and N$_2$, the ratio of the formed N$_2$ column density to the remaining intact NH$_3$ column density was chosen for comparison with observational data reported for molecular clouds \citep{Boogert2015} and comet 67P \citep{Rubin2015, Rubin2019}. Figure~\ref{fig:fig005} shows the derived ratio of $\mathrm{N_2}/\mathrm{NH_3}$ as a function of UV fluence from the photolysis experiments of NH$_3$ ice at 15, 25, and 70~K. Two observational data sets are also shown in Fig.~\ref{fig:fig005} to guide the discussion of astrochemical implications. In a review paper by \citet{Boogert2015}, the median abundance of NH$_3$ ice is reported as 6\% relative to H$_2$O, whereas N$_2$ ice is only constrained by an upper-limit range (i.e., $\lesssim 0.2$--60\% with respect to H$_2$O). Accordingly, the green dashed line shows the lower end of this upper-limit range, corresponding to $\mathrm{N_2}/\mathrm{NH_3}$ $\lesssim 0.03$. The orange dashed line represents the N$_2$/NH$_3$ ratio of $0.13 \pm 0.075$ reported for comet 67P, where NH$_3 = 0.67 \pm 0.20\%$ and N$_2 = 0.089 \pm 0.024\%$ \citep{Rubin2019}. The corresponding UV-fluence ranges shown for protostars and protoplanetary disks are extrapolations of representative cosmic-ray-induced UV photon fluxes reported for UV-irradiated regions in those environments over typical evolutionary timescales.

The experimental results at 15~K show a nearly linear increase across the entire UV-fluence window. The rapid conversion from NH$_3$ to N$_2$ leads to an $\mathrm{N_2}/\mathrm{NH_3}$ ratio reaching the estimated lower end of the upper-limit range inferred for molecular clouds ($\lesssim 0.03$). This indicates that N$_2$ formation directly from NH$_3$ can already account for the lower bound of the N$_2$ abundance in the interstellar ice mantle, even in the absence of N$_2$ gas accretion or in situ N + N $\rightarrow$ N$_2$ reactions. It should be noted that the apparent ice-retained N$_2$ yield at 25 and 70~K remains very low at irradiation doses characteristic of dense clouds.

In the UV fluence regime associated with later star-forming regions ($>1\times10^{18}$~photons~cm$^{-2}$), the experimental data at 15 and 25~K show a further increase in the $\mathrm{N_2}/\mathrm{NH_3}$ ratio to $\sim$0.15 and $\sim$0.11, respectively. These values are comparable to the N$_2$/NH$_3$ = $0.13 \pm 0.075$ ratio measured in the coma of comet 67P, suggesting that prolonged UV photon processing of pure NH$_3$ ice can further enhance the N$_2$ abundance and may, by itself, explain the observed N$_2$ content in comet 67P at UV irradiation fluences $<3.8\times10^{18}$~photons~cm$^{-2}$. This conclusion is consistent with the parallel study of \citet{McKinnon2026}, who showed that UV and electron processing of NH$_3$-bearing interstellar ice analogs can produce N$_2$ at levels relevant to comet 67P. In addition, the present work reports, for the first time, the kinetic evolution of newly formed N$_2$ with respect to NH$_3$ as a function of UV fluence, together with its temperature dependence, during the UV photolysis of pure NH$_3$ ice.

The $\mathrm{N_2}/\mathrm{NH_3}$ ratio at 70~K is consistently the lowest across the entire studied UV-fluence window. An equilibrium value of $\sim$0.04 is reached at a fluence of $\sim3\times10^{18}$~photons~cm$^{-2}$. Although not the primary focus of this work, the photochemical conversion of NH$_3$ ice to N$_2$ at higher temperatures is more relevant to conditions on icy moons in the Solar System, such as Titan and Enceladus, or trans-Neptunian objects (TNOs), such as Charon. Under such environments, solid-state N$_2$ formed via photolysis of NH$_3$ will desorb upon formation, enriching the atmospheres of icy moons or escaping into interplanetary space. Assuming that the difference between the apparent ice-retained N$_2$ yields in the 25 and 70~K irradiation experiments is primarily caused by desorption of formed N$_2$ and some of its precursors (see Sect.~3.2), an upper limit on the apparent N$_2$ loss associated with irradiation of pure NH$_3$ ice, equal to the yeild of gaseous N$_2$, can be estimated. At 70~K, this upper-limit value is $(1.8 \pm 0.4)\times10^{15}$~molecules~cm$^{-2}$ for a total UV fluence of $3.8\times10^{18}$~photons~cm$^{-2}$.

\citet{Gudipati2018} estimate the VUV--UV solar flux reaching Titan to be $\sim1\times10^{14}$~photons~cm$^{-2}$~s$^{-1}$, with a Ly$\alpha$ contribution of $\sim3\times10^{9}$~photons~cm$^{-2}$~s$^{-1}$. It should be noted that this flux refers to Titan's upper atmosphere and not directly to the surface. Therefore, any corresponding UV dose received by NH$_3$ ice at the surface of Titan would be strongly reduced by atmospheric attenuation and would be significantly lower than the dose received by NH$_3$ in atmospheric aerosols. For this reason, the present estimate is more appropriately regarded as relevant to NH$_3$ exposed in Titan's atmospheric aerosols than to NH$_3$ ice at the surface. The length of a Titan day is about $1.4\times10^{6}$~s, corresponding to a fluence of $\sim1.4\times10^{20}$~photons~cm$^{-2}$. This results in a yield of gaseous N$_2$ of $\sim6.6\times10^{16}$~molecules~cm$^{-2}$ over one Titan day, or $\sim1.5\times10^{18}$~molecules~cm$^{-2}$ over one Earth year. The N$_2$ column density on Titan's surface can be estimated based on the mass of Titan's atmosphere ($\sim9.1\times10^{18}$~kg; \citealt{Vuitton2023}) and its surface area ($\sim8.3\times10^{7}$~km$^{2}$). Assuming a 95\% fraction of N$_2$ in the atmosphere and using the molecular mass of N$_2$, a column density of $\sim2.2\times10^{26}$~molecules~cm$^{-2}$ is derived. Under idealized conditions in which NH$_3$ ice constitutes a major component of the aerosols in the atmosphere, the estimated atmospheric N$_2$ column density could be produced within $\sim1.5\times10^{8}$~yr, which is below the age of the Solar System. It should be noted, however, that this estimate represents an upper-limit contribution from NH$_3$ ice processing, while gas-phase ion--neutral reactions in Titan's atmosphere also provide efficient pathways for N$_2$ production. Further insights into the contribution of NH$_3$ ice energetic processing to the enrichment of Titan's atmosphere may be obtained through analysis of the corresponding isotopic fractionation.

\section{Conclusions}
The main experimental findings are summarized below:
\begin{itemize}

\item Laboratory UV photolysis of pure NH$_3$ ice demonstrates efficient formation of nitrogen-bearing products, including N$_2$ and N$_2$H$_4$. For the first time, the formation abundances of both N$_2$ and N$_2$H$_4$ are reported simultaneously as a function of applied UV photon fluence for ice temperatures of 15, 25, and 70~K.

\item Initial formation yields were obtained by linear fitting of the product kinetic evolutions up to a UV fluence of $1\times10^{18}$~photons~cm$^{-2}$ for N$_2$ and $2\times10^{18}$~photons~cm$^{-2}$ for N$_2$H$_4$, corresponding to the typical UV fluence accumulated in dense clouds over $\sim10^{6}$~yr, and are reported for ice temperatures of 15, 25, and 70~K. For N$_2$, these values represent apparent ice-retained formation yields in the solid state.

\item The ratio of N$_2$ to NH$_3$ ice column densities was calculated as a function of UV fluence and compared with observational constraints for dense molecular clouds ($\lesssim0.03$, the lower end of the upper-limit range) and comet 67P ($\sim0.13$). Irradiation of pure NH$_3$ ice is shown to reproduce the cometary $\mathrm{N_2}/\mathrm{NH_3}$ ratio at UV fluences up to $3.8\times10^{18}$~photons~cm$^{-2}$, characteristic of protoplanetary disk evolutionary stages.
\end{itemize}

Finally, it should be noted that the experiments presented here focus on pure NH$_3$ ice under controlled laboratory conditions, whereas astrophysical ices are chemically heterogeneous, and the presence of additional ice constituents may therefore influence reaction pathways and apparent solid-state product yields of N$_2$ and N$_2$H$_4$. Within this context, the upper limit on the apparent N$_2$ loss associated with UV irradiation of pure NH$_3$ ice at 70~K is determined to be $(1.8 \pm 0.4)\times10^{15}$~molecules~cm$^{-2}$ for a total UV fluence of $3.8\times10^{18}$~photons~cm$^{-2}$; together with the solid-state N$_2$ and N$_2$H$_4$ yields, this provides useful constraints for future investigations of icy-moon atmospheres by missions such as ESA's JUICE and NASA's Dragonfly.

\begin{acknowledgements}
This work has been supported by the Danish National Research Foundation
through the Center of Excellence InterCat (Grant Agreement No.~DNRF150).
It has also been funded by the Dutch Astrochemistry Network II (DANII)
and NOVA, the Netherlands Research School for Astronomy.
GF acknowledges support from the Xinjiang Tianchi Talent Program (2024).
\end{acknowledgements}

\bibliographystyle{aa}
\bibliography{Ammonia}

\begin{appendix}

\section{Mass spectrometry validation of detected species}

To confirm the molecular assignment of NH$_3$ and N$_2$H$_4$ in our experiments, we compared the mass spectra obtained in this work with reference spectra from the NIST Chemistry WebBook database. The comparison shows good agreement in the relative fragmentation patterns and the main diagnostic peaks, supporting the identification of the desorbed products as NH$_3$ and N$_2$H$_4$. Minor deviations in peak intensities can be attributed to differences in instrumental response and ionization conditions between our TOF mass spectrometer and the mass spectrometers used for the NIST reference data sets.

\begin{figure}[h]
\centering
\includegraphics[width=0.65\linewidth]{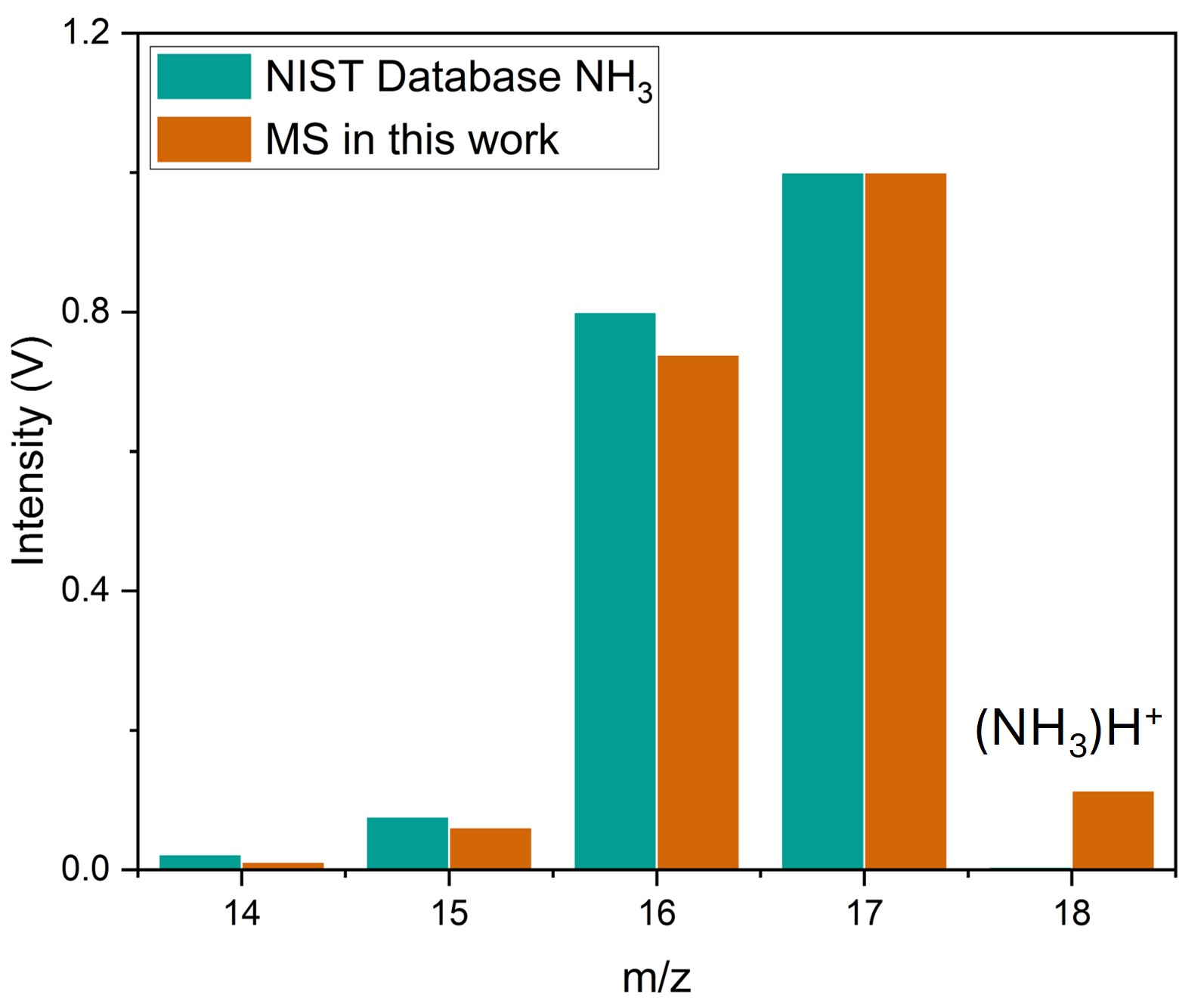}
\caption{Comparison of the NH$_3$ mass spectrum obtained in this work (orange) with the NIST database spectrum (blue). The main peaks at $m/z = 16$ and $17$, corresponding to NH$_2$$^{+}$ and NH$_3$$^{+}$, are well reproduced, confirming the identification of NH$_3$.}
\label{fig:fig0021}
\end{figure}

\begin{figure}[h]
\centering
\includegraphics[width=0.65\linewidth]{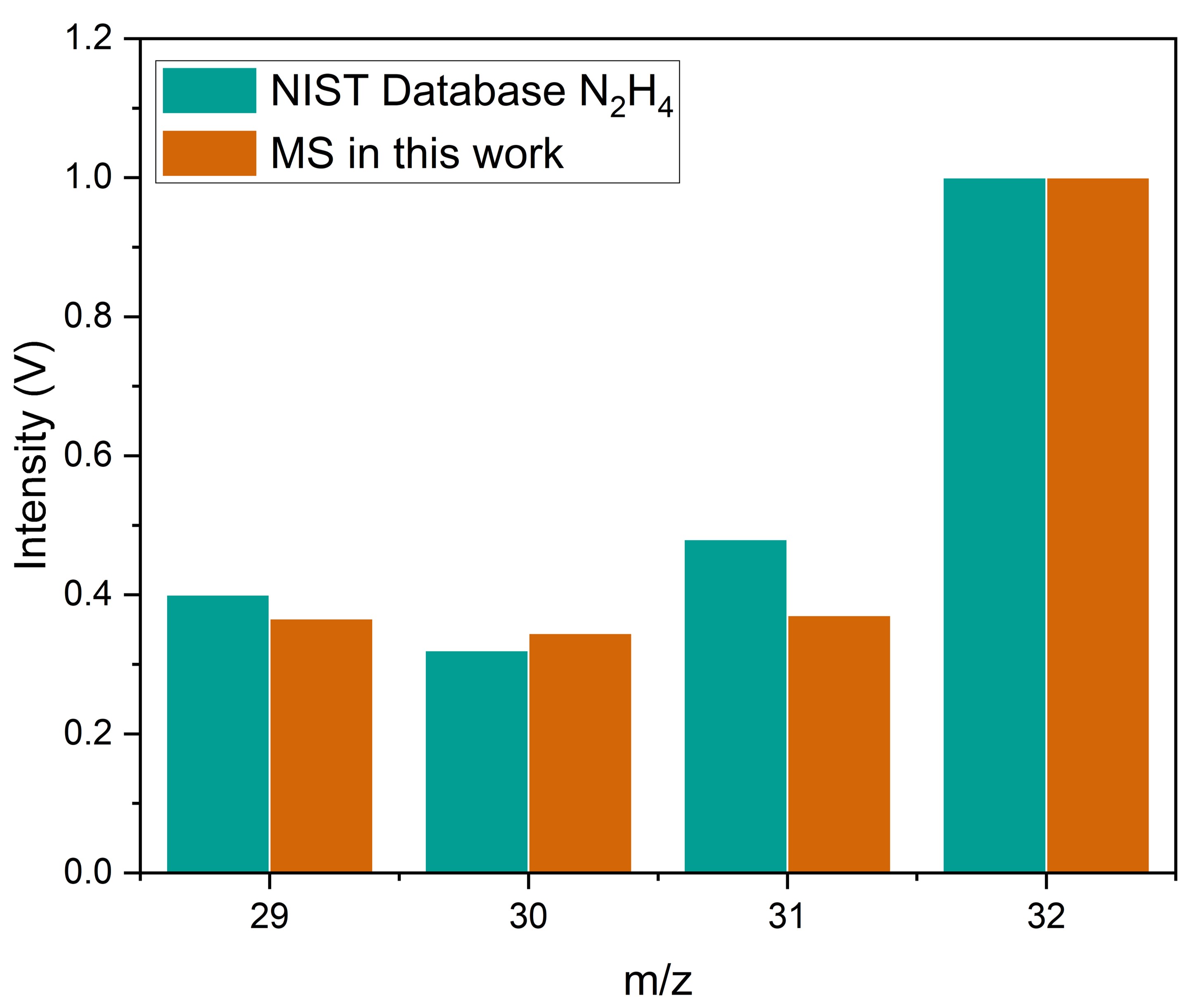}
\caption{Comparison of the N$_2$H$_4$ mass spectrum obtained in this work (orange) with the NIST database spectrum (blue). The fragmentation pattern with major contributions at $m/z = 29$--$32$ is consistent with the NIST reference, confirming the assignment of N$_2$H$_4$.}
\label{fig:fig0022}
\end{figure}

\end{appendix}

\end{document}